\documentclass[final,3p,times]{elsarticle}

\usepackage{graphicx}
\usepackage{amssymb}
\usepackage{amsmath}
\usepackage{lineno}
\usepackage{color}
\usepackage{url}
\usepackage{setspace}
\newcommand*\patchAmsMathEnvironmentForLineno[1]{%
  \expandafter\let\csname old#1\expandafter\endcsname\csname #1\endcsname
  \expandafter\let\csname oldend#1\expandafter\endcsname\csname end#1\endcsname
  \renewenvironment{#1}%
  {\linenomath\csname old#1\endcsname}%
  {\csname oldend#1\endcsname\endlinenomath}}% 
\newcommand*\patchBothAmsMathEnvironmentsForLineno[1]{%
  \patchAmsMathEnvironmentForLineno{#1}%
  \patchAmsMathEnvironmentForLineno{#1*}}%
\AtBeginDocument{%
  \patchBothAmsMathEnvironmentsForLineno{equation}%
  \patchBothAmsMathEnvironmentsForLineno{align}%
  \patchBothAmsMathEnvironmentsForLineno{flalign}%
  \patchBothAmsMathEnvironmentsForLineno{alignat}%
  \patchBothAmsMathEnvironmentsForLineno{gather}%
  \patchBothAmsMathEnvironmentsForLineno{multline}%
}
\newcommand*{\PDEVUV} {PDE_{\rm{VUV}}}
\newcommand*{\PDEvis} {PDE_{\rm{vis}}}

\journal{Journal Name}

\begin{document}

\begin{frontmatter}


    \title{Study on degradation of VUV-sensitivity of MPPC for liquid xenon scintillation detector by radiation damage in MEG II experiment}

    \author[label1]{K. Ieki}
    \author[label1]{T. Iwamoto}
    \author[label1]{S. Kobayashi \corref{cor0}} \cortext[cor0]{Corresponding author} \ead{satoruk@icepp.s.u-tokyo.ac.jp}
    \author[label1]{Toshinori Mori}
    \author[label1]{S. Ogawa}
    \author[label1]{R. Onda}
    \author[label1]{W. Ootani}
    \author[label1]{K. Shimada}
    \author[label1]{K. Toyoda}
    \address[label1]{International Center for Elementary Particle Physics (ICEPP), The University of Tokyo, Tokyo, 113-0033 Japan}

    % \address{The University of Tokyo}

    \begin{abstract}
        In the MEG II experiment, the liquid xenon gamma-ray detector uses Multi-Pixel Photon Counters (MPPC) which are sensitive to vacuum ultraviolet (VUV) light under a high-intensity muon beam environment.
        In the commissioning phase of the detector with the beam, a significant degradation in the photon detection efficiency (PDE) for VUV light was found, while the degradation in the PDE for visible light was much less significant.
        This implies that the radiation damage is localized to the surface of the MPPC where incoming VUV photons create electron-hole pairs.
        It was also found that the PDE can recover to the original level by thermal annealing.
        % As a countermeasure for the observed radiation damage, thermal annealing was tested, and recovery of the VUV PDE up to the initial level was achieved.
    \end{abstract}


\end{frontmatter}

%\tableofcontents
% \linenumbers

\section{Introduction}\label{sec:Intro}
% SiPM, VUV-sensitive MPPC
As liquid noble gas scintillation detectors are employed in different particle physics experiments, demand for photo-sensors sensitive to the liquid noble gas scintillation in the VUV range is increasing.

A VUV-sensitive multi-pixel photon counter (MPPC, one of SiPM families) was developed for the liquid xenon (LXe) detector of the MEG II experiment in collaboration with Hamamatsu Photonics K.K. (S10943-4372)~\cite{Ieki2019}.
A high PDE of 20$\pm$2\%~\cite{Ieki2019} was achieved for the LXe scintillation light peaked at $174.8\pm 0.1 (\mathrm{stat.})\pm0.1(\mathrm{syst.})\,\rm{nm}$~\cite{LXeFujii}.
The detection mechanism of VUV photons is slightly different from that of visible photons.
A visible photon can directly reach the signal amplification region of the MPPC, and trigger an avalanche.
On the other hand, a VUV photon is absorbed in the vicinity of the surface of the silicon chip due to its short attenuation length,
and the generated charge carrier must drift and diffuse to the amplification region to trigger an avalanche.
Since the VUV-sensitive MPPC is a new photosensor, there is little knowledge and experience on the radiation hardness.
Its radiation hardness and long-term stability are important for experiments and applications in the future.

The LXe gamma-ray detector for the MEG~II experiment is the first large-scale detector that uses the VUV-sensitive MPPCs~\cite{Baldini2018}.
It is designed to detect a $52.8\,\rm{MeV}$ gamma-ray originating from a charged lepton flavor violating decay of a muon, $\mu^{+}\to e^{+}\gamma$.
The high granularity and uniformity of the scintillation readout realized by the MPPCs lead to a good position and energy resolution for the gamma-ray.

In the MEG~II experiment, a high-intensity muon beam (up to $7\times 10^{7}\,\rm{s^{-1}}$) is injected into a muon stopping target, which is located $\sim 60~\rm{cm}$ away from the LXe gamma-ray detector.
The MPPCs are exposed to gamma-rays, neutrons, and VUV photons; only neutral particles reach the LXe detector because a magnetic field is applied around the stopping target.

% Explanation of this article
This article reports a study on the radiation damage of VUV-sensitive MPPCs in the MEG II LXe detector under the high-intensity muon beam environment.
This paper is organized as follows.
In Section~\ref{sec:Method}, the design of the LXe detector and the dose level with the muon beam are described.
In Section~\ref{PDEDegradation}, degradation of the PDE under the high-intensity muon beam environment is reported.
In Section~\ref{Annealing}, recovery of the radiation damage by thermal annealing is described.
Section~\ref{Conclusion} concludes this article.

\section{MEG II liquid xenon gamma-ray detector}\label{sec:Method}
% The irradiation test was performed by operating the LXe detector for MEG~II experiment under a high-intensity muon beam environment.

\subsection{Liquid xenon gamma-ray detector}
\begin{figure}[htb]
    \centering
    \includegraphics[width=0.9\hsize]{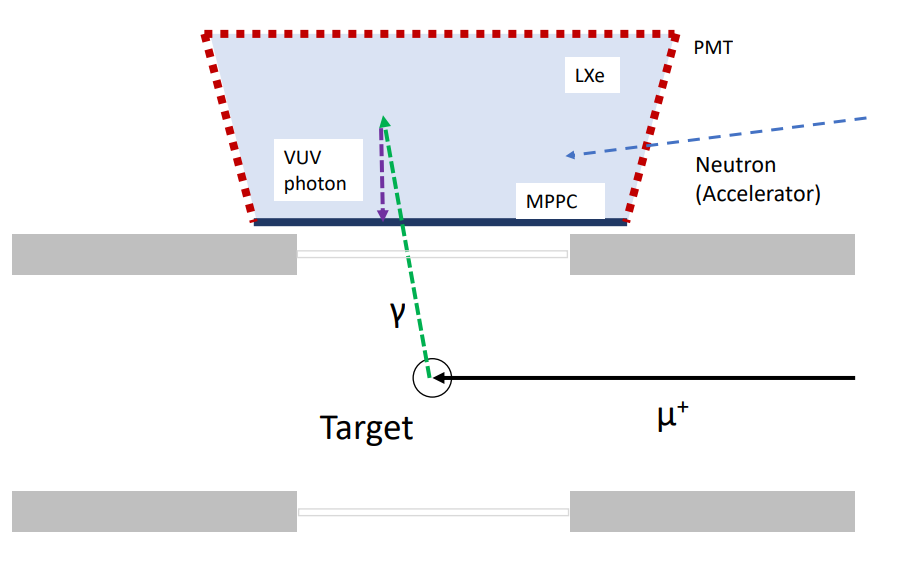}
    \caption{Radiation environment from the muon beam at the MEG~II experimental area.}
    \label{fig:RadType}
\end{figure}

The LXe detector is designed to measure the gamma-ray from $\mu^{+}\to e^{+}\gamma$ with good position, timing, and energy resolutions.
LXe has several advantages for this purpose as a scintillator, such as a high light yield (the mean excitation energy, $W = 21.6\,\rm{eV}$/photon), fast decay time ($45\,\rm{ns}$), and high stopping power.
A $900\,\mathrm{\ell}$ LXe volume is surrounded by two different types of VUV-sensitive photosensors which are immersed in LXe and are working at 165~K.
The gamma-ray entrance face ($0.92\,\rm{m}^{2}$) of the volume is covered by 4,092 VUV-sensitive MPPCs with the size of $12\times12\,\mathrm{mm^2}$ in a regular grid structure.
The other surfaces of the detector are covered by 668 VUV-sensitive 2-inch PMTs in diameter (Fig.~\ref{fig:RadType}).

\begin{figure}[htb]
    \centering
    \includegraphics[width=\hsize]{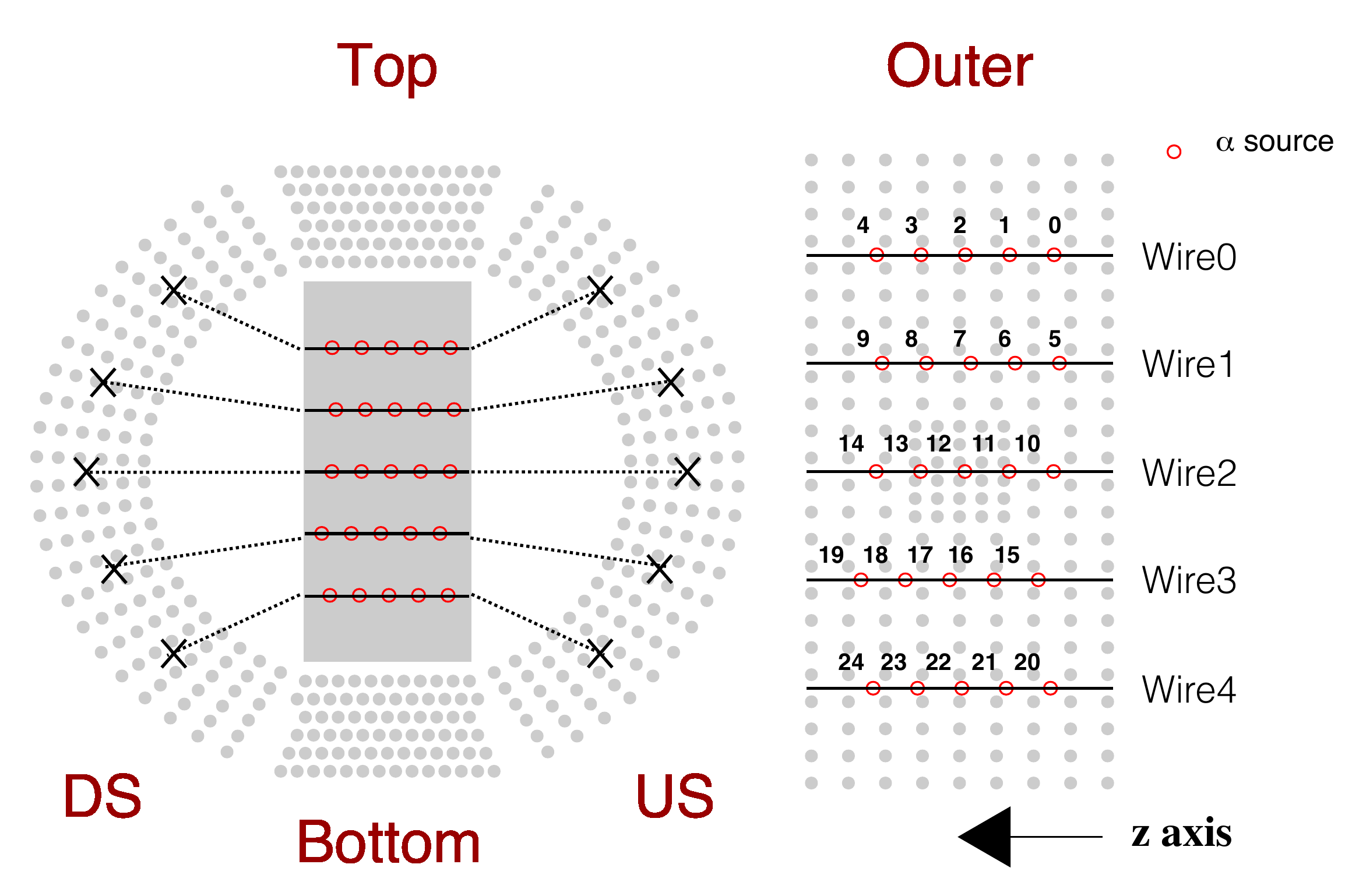}
    \caption{Position of 25 $\times$ $^{241}$Am sources on five tungsten wires. Red circles on the entrance face and the outer face illustrate the projected position of the sources. Z axis is the same as the beam axis in the MEG II experiment.}
    \label{fig:alphasourcegeometry}
\end{figure}

% Alpha source, LED
The detector is equipped with two types of light sources to calibrate the photosensors.
% Alpha
Twenty-five $^{241}$Am spot $\alpha$-ray sources are installed to measure the PDE of the photosensors for the LXe scintillation light as shown in Fig.~\ref{fig:alphasourcegeometry}. They emit alpha-rays at $5.485\,\mathrm{MeV}$  (84.5\%) and $5.443\,\mathrm{MeV}$ (13.0\%) which produce VUV scitntillation light.
A 1.5$\,\mathrm{\mu m}$-thick gold foil is wound around a thin gold-plated tungsten wire with a diameter of 100$\,\mathrm{\mu m}$~\cite{BALDINI2006589}.
%Each source is surrounded by a thin gold foil and wound around a thin gold-plated tungsten wire with a diameter of 100$\,\mathrm{\mu m}$~\cite{BALDINI2006589}.
The activity of each source is $\sim200\,\rm{Bq}$, which is small enough not to interfere with other measurements.
% LED
The other light source is blue LEDs (Toyoda Gosei E1L49-3B1A-02) mounted on the lateral faces of the detector.
The blue light from the LEDs ($\lambda \sim 460~\rm{nm}$) can be used to monitor the gain and the PDE of the photosensors for the visible light.
They also allow us to measure correlated noises of the MPPCs as described in \ref{LEDCalib}.

\begin{figure}[htb]
    \centering
    \includegraphics[width=\hsize]{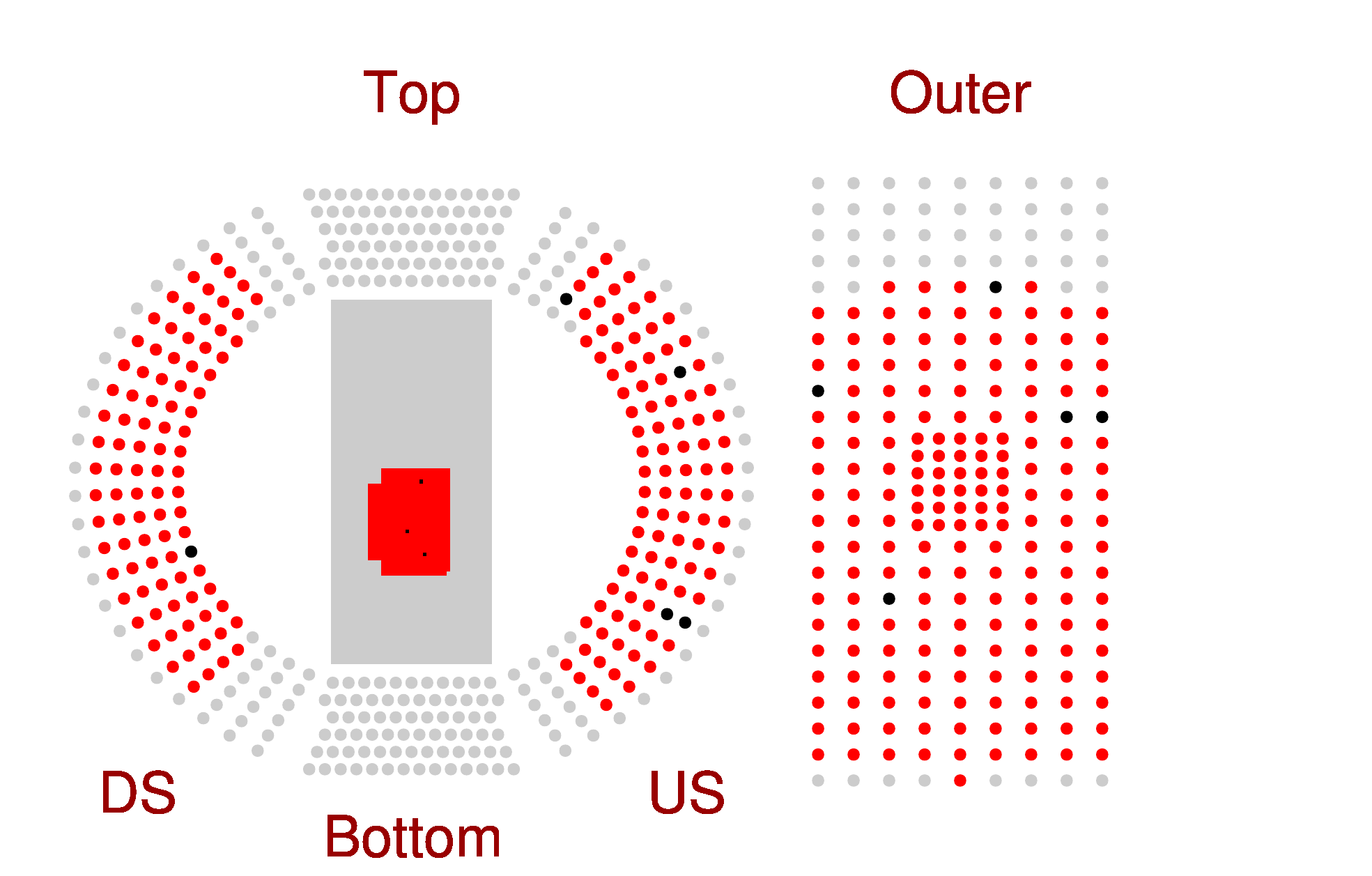}
    \caption{Channels to which readout electronics were assigned in 2018 and 2019 (red), no readout electronics were assigned (grey), and dead channels (black).}
    \label{fig:pmmap}
\end{figure}

% Electronics
The signals of the photosensors are transmitted through coaxial cables to the readout electronics (WaveDREAM modules~\cite{GALLI2019399}), where the waveforms are recorded.
The measurement described in this article was carried out in a preliminary phase of the experiment when the total number of readout channels was limited to about 1,000.
Thus, 640 MPPCs out of 4,092 and 360 PMTs out of 668 shown in Fig.~\ref{fig:pmmap} were read out.
The bias voltage of the MPPC was applied only to the read-out channels.

\subsection{Exposure to radiation}
% \subsection{Beam test}
This paper discusses the radiation damage of the VUV-sensitive MPPCs based on the observation during the commissioning of the LXe detector with the muon beam from 2017 to 2019.
A high-intensity muon beam up to $7\times 10^{7}\,\rm{s^{-1}}$ was used for 220~hours in 2017, 150~hours in 2018, and 230~hours in 2019.

Fig.~\ref{fig:RadType} illustrates the radiation environment during the beamtime. 
The MPPCs were irradiated with gamma-rays, neutrons, and VUV photons.
The radiation dose in 2019 summarized in Table.~\ref{tab:exprad} is estimated as follows.

% Gamma-ray
Gamma-rays come from radiative muon decays ($\mu\to e\nu \bar{\nu}\gamma$) and pair annihilation of positrons from muon decays.
Some of them reach the detector and can cause an ionizing energy loss in the MPPCs.
The total dose from the gamma-ray in the beam time is estimated to be $1\times10^{-4}\,\mathrm{Gy}$ based on a Geant4 simulation.

% Neutron
Neutrons come from the hadronic interaction of the primary proton beam with the pion production target, which is located 20 m away upstream of the muon beam-line from the MEG detectors.
The total neutron fluence in the 2019 run is estimated to be less than $3 \times 10^{6}\, n/\rm{cm}^{2}$ from the measured neutron flux in the experimental area.

% VUV photon
In addition, the MPPCs are exposed to the VUV photons from the LXe scintillation.
The photon fluence during the beamtime in 2019 is estimated to be $6 \times 10^{10}\,\rm{photon/mm^{2}}$, based on the MPPC current measurement and Geant4 simulation. The MPPC current is the product of the photon fluence, PDE, and gain. The PDE and gain were measured with the calibration measurement.
Geant4 simulation gives the number of incident photons with a beam rate.

% Move to Discussion?
It is notable that these are far smaller than the dose level at which an increase of the leakage current is reported for standard MPPCs~\cite{Matsubara:20083N}\cite{Matsumura:20081D}.
% Typical number of radiation dose
For example, Ref.~\cite{Matsubara:20083N} reported an increase of the leakage current after 240~Gy of $^{60}\rm{Co}$ gamma-ray irradiation.

\begin{table}[htb]
    \centering
    \caption{Exposure to radiation in 2019 run}
    \label{tab:exprad}
    \begin{tabular}{ll}
        \hline
        Particle   & Dose / Fluence                                                   \\
        \hline
        Gamma-ray  & $1 \times10^{-4}\,\mathrm{Gy}$                                   \\
        Neutron    & $3 \times 10^{6}\, \rm{cm}^{-2}$ ($1\,\mathrm{MeV}$ equivalent) \\
        VUV photon & $6 \times 10^{10}\, \rm{mm^{-2}}$                          \\
        \hline
    \end{tabular}
\end{table}
% Commissioning

\section{PDE degradation}\label{PDEDegradation}
\subsection{PDE measurement}

% Measurement principle
The PDE for VUV light $\PDEVUV$ was measured by using the xenon scintillation light from the energy deposit of the alpha particle.
The PDE is defined as a ratio of the number of detected photoelectrons ($\bar{N}_{\text{pe,obs}}$) to the expected number of incident photons ($\bar{N}_{\text{ph,expected}}$);

\begin{equation}\label{eq:PDE}
    PDE = \bar{N}_{\text{pe,obs}}/\bar{N}_{\text{ph,expected}} .
\end{equation}
The $\bar{N}_{\text{pe,obs}}$ is calculated from the integrated charge of the waveform $\bar{Q}_{\text{obs}}$, the gain $G$, and the excess charge factor (ECF, $F_{\text{EC}}$) of the photosensor;
\begin{equation}\label{eq:Npe}
    \bar{N}_{\text{pe,obs}} = \frac{\bar{Q}_{\text{obs}}}{G \times F_{\text{EC}}} .
\end{equation}
Here the $F_{\text{EC}}$ represents the amplification of the charge by correlated noises such as cross-talk and after-pulsing.
The gain and the ECF are measured using LEDs as described in \ref{LEDCalib}.

The $\bar{N}_{\text{ph,expected}}$ is estimated by using a Geant4-based Monte-Carlo simulation.
The simulation calculates the propagation of scintillation photons from the energy deposit of the alpha ray based on the parameters summarized in Table.~\ref{tab:MCparams}.
The absorption length of LXe in MC simulation is set to 400 cm, which is much longer than the detector size, because no finite absorption was observed in data.

The PDE for visible light $\PDEvis$ was measured with the blue LED light.
The absolute $\PDEvis$ value cannot be estimated because it is difficult to estimate the expected number of photons and thus only the time variation was measured.

% Systematics of the PDE measurement
Systematics of the absolute $\PDEVUV$ value come mainly from the error of the scintillation light yield for $\alpha$-rays (10\%~\cite{Doke1999}) and the uncertainty in the estimation of the effect of the reflection on the surface of the inner wall of the detector (5\%).

\begin{table}[htb]
    \centering
    \caption{Parameters in MC simulation}
    \label{tab:MCparams}
    \begin{tabular}{ll}
        \hline
        Parameter                  & Value                                \\
        \hline
        Absorption length          & $400\,\mathrm{cm}$                   \\
        Rayleigh scattering length & $45\,\mathrm{cm}$~\cite{1657967} \\
        $W$-value for $\alpha$-ray   & $19.6\,\mathrm{eV}$~\cite{1999NIMPA.420...62D}   \\
        Geant4 version             & 10.4.p03                             \\
        \hline
    \end{tabular}
\end{table}

\subsection{Degradation of PDE}
\begin{figure}[htb]
    \centering
    \includegraphics[width=\hsize]{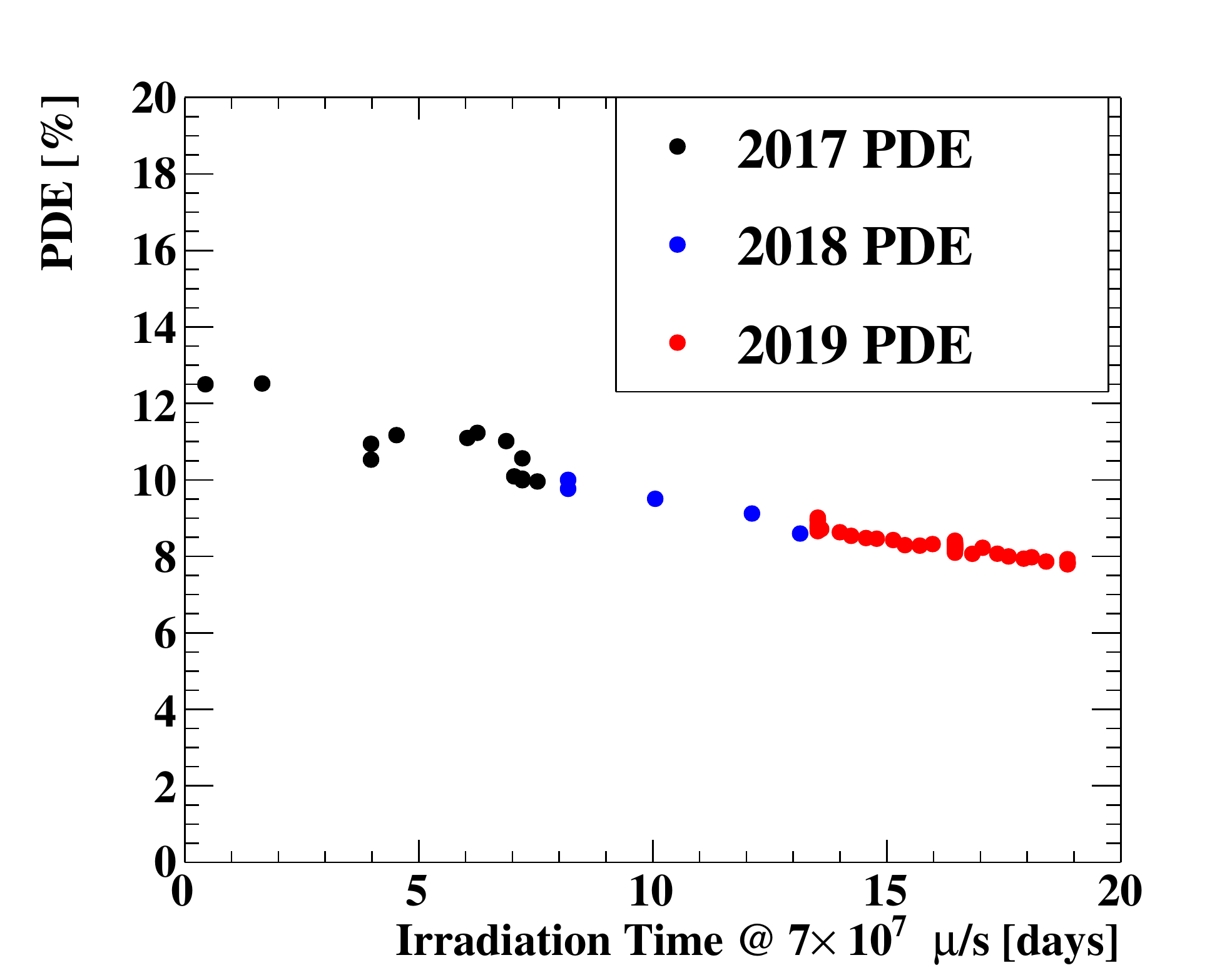}
    \caption{MPPC PDE time variation from 2017 to 2019. The horizontal axis is the irradiation time, which is normalized with $7\times 10^{7}~/s$ at the stopping target. The PDE is averaged over readout MPPCs.}
    \label{fig:PDEhist}
\end{figure}

%2018
Fig.~\ref{fig:PDEhist} shows the measured $\PDEVUV$ of the MPPC during the beam time in 2017, 2018, and 2019.
The radiation damage under a high-intensity muon beam was strongly suspected because no degradation had been observed within the absolute precision of 1\% without the muon beam for 300 days between 2017 and 2018 runs.
These observations motivated frequent PDE measurements with the muon beam in 2019.

\begin{figure}[htb]
    \centering
    \includegraphics[width=\hsize]{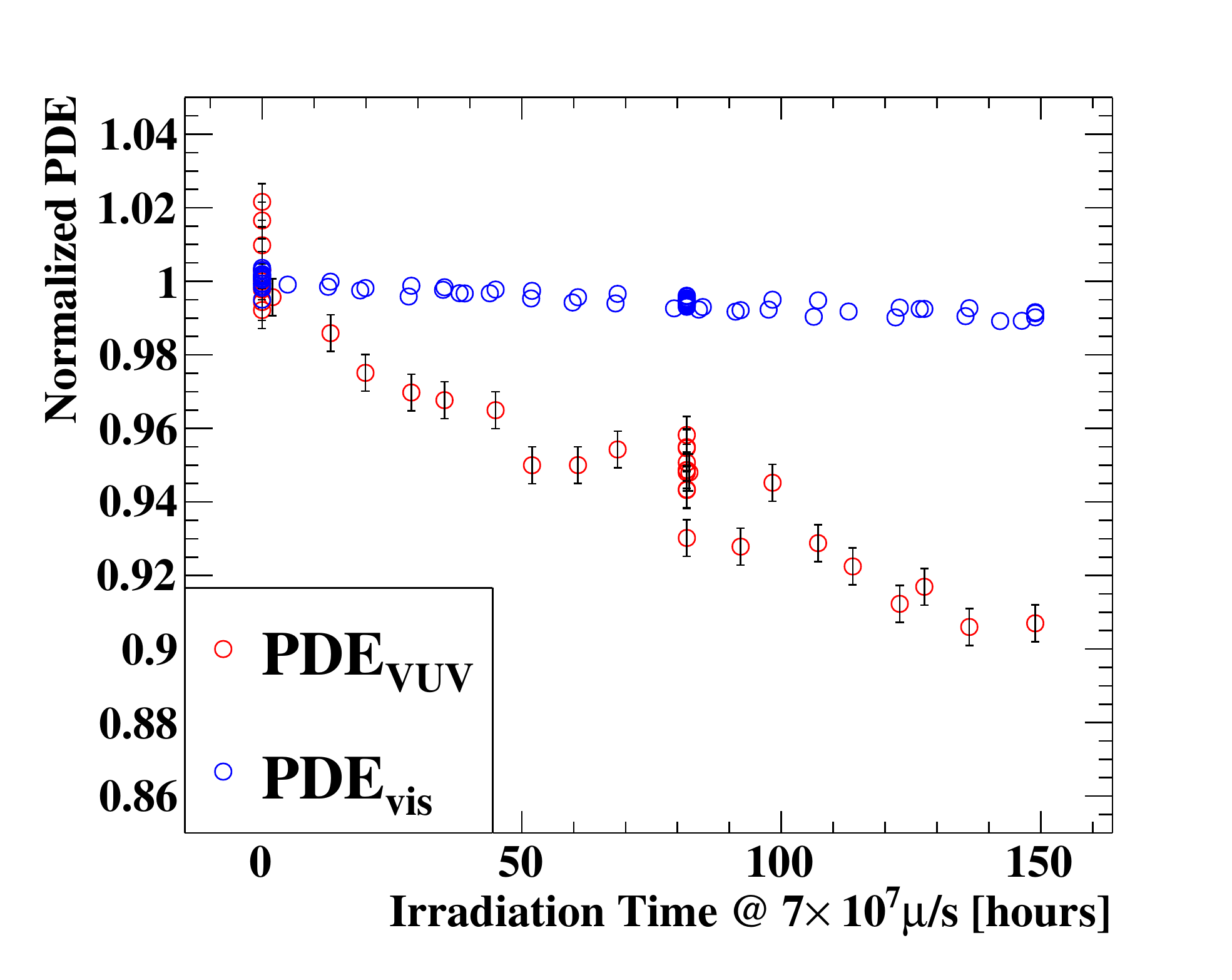}
    \caption{Time variation of the average PDE for VUV light ($\PDEVUV$, red) and for the visible light ($\PDEvis$, blue). Both graphs are normalized at the beginning of the beam time. $\PDEVUV$ gradually decreased by 10\% and $\PDEvis$ decreased by 1\%. The error bars represent only the statistical uncertainty.}
    \label{fig:2019VUV}
\end{figure}

In 2019 run, the PDE was measured every 11 hours during the beam time; a set of the irradiation (8~hours) and the PDE measurement without irradiation (3~hours) was repeated.
% Relative PDE deterioration
Fig.~\ref{fig:2019VUV} shows the time variation of the $\PDEVUV$ and $\PDEvis$ averaged over the readout channels during the 2019 beam time.
The $\PDEVUV$ gradually decreased by 10\% in 150~hours of intermittent muon beam usage.
The $\PDEvis$ also decreased by 1\% during the beam time, and the decrease rate was one order of magnitude smaller than that of the $\PDEVUV$.
As we discuss in Section~\ref{surfacedamage}, this wavelength dependence of the degradation implies that it is caused by surface damage.

\begin{figure}[htb]
    \centering
    \includegraphics[width=\hsize]{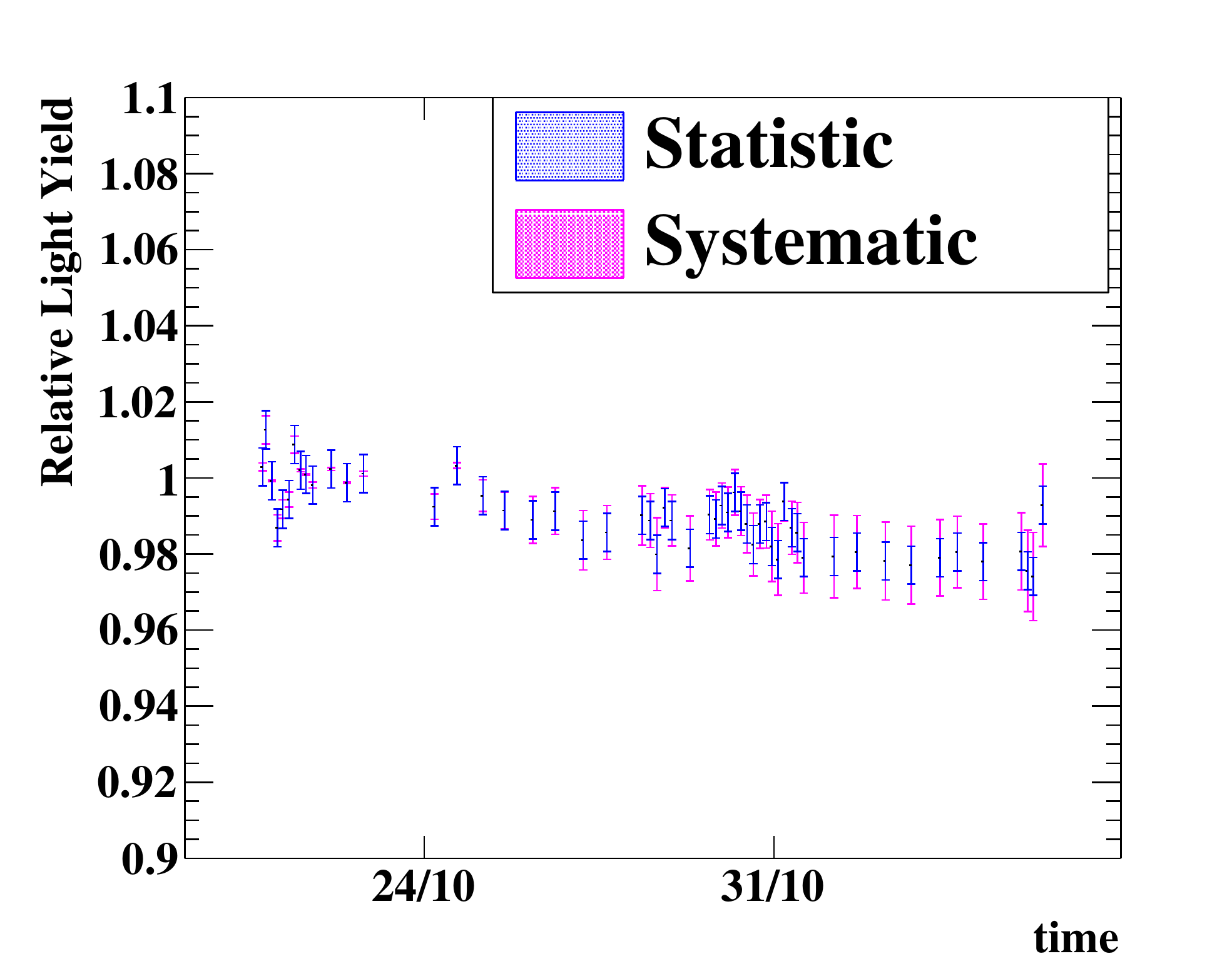}
    \caption{Time variation of light yield of LXe in 2019. Normalized at the beginning of the beam time. See the text about the detail of the systematic uncertainty.}
    \label{fig:PMTPDE}
\end{figure}

% Uncertainty
Impurities in the LXe can cause a systematic error in $\PDEVUV$ measurement because they reduce the emission of the scintillation photons or they absorb the scintillation photons during the propagation and reduce the number of detected photons by the MPPCs.
The effect of impurities was monitored by using the PMTs. Both the detected number of visible photons from LEDs and that of VUV photons from the $\alpha$ sources were monitored by PMTs during the irradiation.
It was found that the number of the VUV photons decreased by 2--4\% compared with that of the visible photons.
This can be caused by either degradation of the LXe purity or degradation of quantum efficiency of the PMT for the VUV light.
Thus, the degradation of the light yield or of the quantum efficiency of the PMT in this period was estimated to be 1--3\% as shown in Fig.~\ref{fig:PMTPDE},
by including both possibilities as a systematic uncertainty.
In any case, the observed PDE degradation of MPPC can not be explained by this instability.

The PDE at the beginning of the first beam time in 2017 was 13$\pm$1\%, which is significantly lower than the value measured in a laboratory test of 20$\pm$2\%.
The radiation damage can explain the low PDE at the beginning of the 2017 run because the MPPCs were exposed to a non-negligible amount of radiation during a beam tuning period before the 2017 run when the PDE monitoring had not started yet.

\subsection{Surface damage}\label{surfacedamage}
The degradation of the PDE for the VUV light ($\lambda = 175\,\rm{nm}$) is much more significant compared to that for the visible light ($\lambda = 460\,\rm{nm}$).
This wavelength dependence can be explained if the damage is localized to the surface of the MPPC because of the much shorter attenuation length of silicon for the VUV light ($\sim 6\,\rm{nm}$) compared to that for the visible light ($\sim 280\,\rm{nm}$).

% Mechanism
\begin{figure}
    \centering
    \includegraphics[width=\hsize]{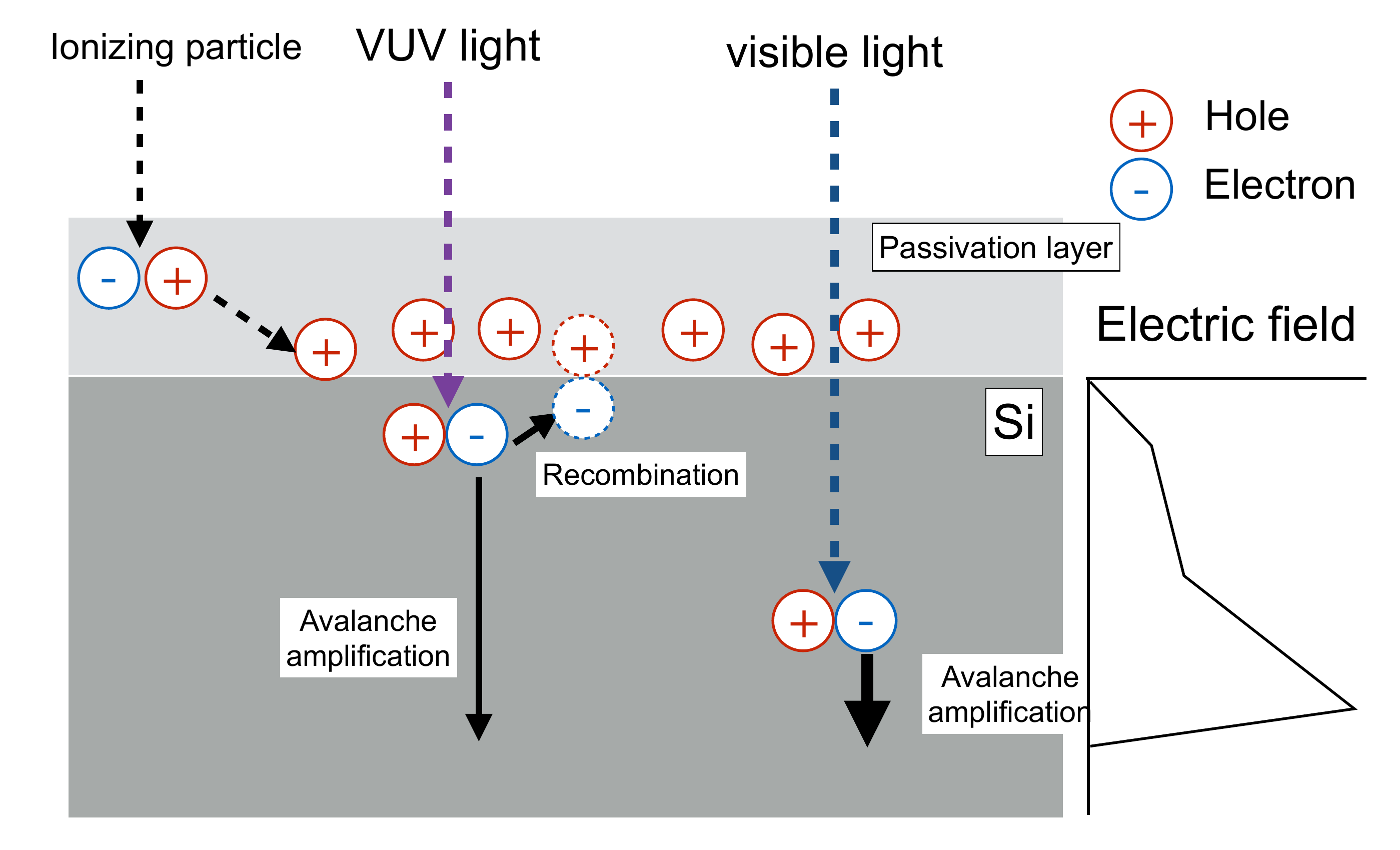}
    \caption{Schematics of surface damage. Radiation can generate electron-hole pairs in the passivation layer and a fraction of holes are trapped in the layer (left). The remaining holes reduce the electric field around the interface and enhance recombination (middle). The effect on visible light is much smaller due to long attenuation length (right)}
    \label{fig:SurfaceDamage}
\end{figure}

The mechanism of the less PDE can be due to enhanced recombination close to the surface of the silicon as explained in Fig.~\ref{fig:SurfaceDamage}.
Firstly, electron-hole pairs are generated in the passivation layer (e.g.$\rm{SiO_{2}}$) by incoming radiation,
and some of them diffuse in the passivation layer.
While most electrons leave the passivation layer because of their high mobility and low trapping probability, some holes remain.
These holes can be trapped by the defects in the passivation layer or at the interface between the silicon layer and the passivation layer.
This hole trapping results in an increase of recombination probability in the vicinity of the interface, where the electron-hole pairs are generated by the VUV light because generated electrons are attracted to the trapped holes.
The recombination around the interface prevents the electrons from reaching the amplification region, and the detection efficiency for the VUV light decreases.
In comparison, visible light is less affected to this effect because it can penetrate deep into the silicon layer and thus directly reach the amplification region.
% Annealing
If this hypothesis is correct, thermal annealing has the potential for the recovery of the PDE.
As in Ref.~\cite{CCD}, the thermal excitation can detrap the holes in the interface state and restores the interface states so that the electrons from electron-hole pairs generated in the vicinity of the interface can reach the amplification region again.

\section{PDE recovery by thermal annealing}\label{Annealing}

\begin{figure}
    \centering
    \includegraphics[width=\hsize]{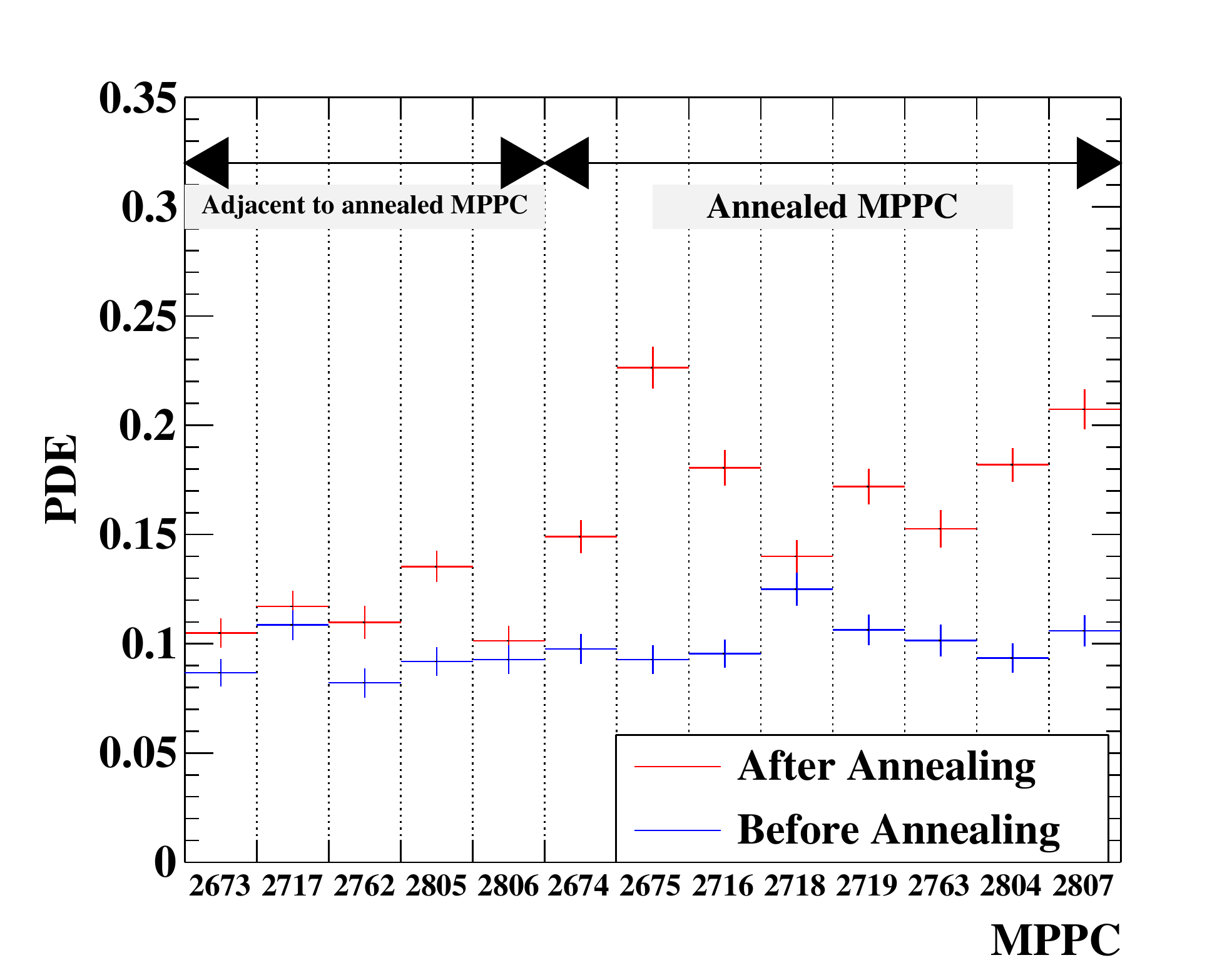}
    \caption{$\PDEVUV$ of MPPCs before the annealing (blue) and after (red). $\PDEVUV$ for MPPCs adjacent to the annealed ones and $\PDEVUV$ for the annealed MPPCs are shown.
    The $\PDEVUV$ was significantly recovered by thermal annealing, and the $\PDEVUV$ of neighboring MPPCs was also recovered by the heat transmitted from the annealed MPPCs.}
    \label{fig:annealing_2020}
\end{figure}

\begin{figure}
    \centering
    \includegraphics[width=\hsize]{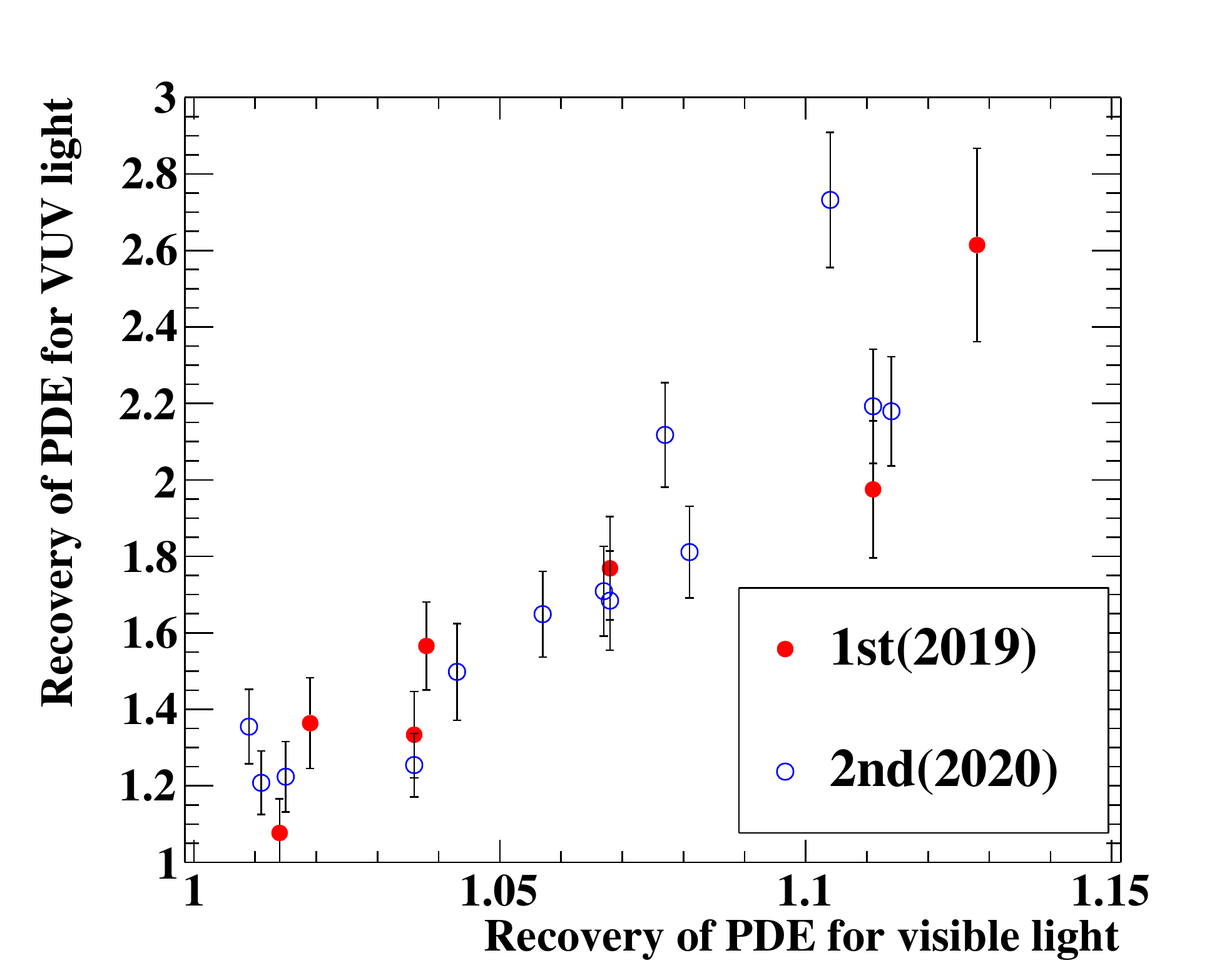}
    \caption{Correlation between the recovery of $\PDEVUV$ and $\PDEvis$. }
    \label{fig:annealing_visvuvcor}
\end{figure}

Thermal annealing was tested for several MPPCs in the LXe detector.
The temperature of the MPPCs was raised by the Joule heat which is generated by photo-electric current by supplying high reverse bias voltage and continuous illumination of intense LED light to MPPCs.
Liquid xenon was transferred to a storage tank during the annealing process, and gaseous xenon at room temperature was filled in the detector.

% Lab test
One concern was heat damage to surrounding material. 
The MPPCs are mounted on printed circuit boards (PCBs), and the PCBs are mounted on carbon frame reinforced plastics (CFRP) support structure. The CFRP can be damaged when it is heated over 45$^{\circ}$C which is the glass-transition temperature of the resin in the CFRP.
To avoid this damage, a test was carried out in a laboratory setup.
The same MPPC and PCB that are used in the actual detector are prepared to reproduce the setup in the LXe detector, and it was illuminated with a room light with a bias voltage applied. Its temperature was measured by a thermal imaging camera (FLIR E50).
It was found that the MPPCs can be heated to 62$^{\circ}$C with the bias voltage of $65\,\mathrm{V}$, which corresponds to the overvoltage of 12~V.
It is much higher than the normal operating over voltage of 7~V, and the induced current of $20\,\mathrm{mA}$ while keeping the backside of the PCB to 39$^{\circ}$C.
The basic properties of the MPPCs were measured before and after the 60 hours of the thermal stress. While the gain and ECF did not change, the dark noise at over-voltage of $3.5\,\mathrm{V}$ was reduced by $\sim$15\%.

% XEC annealing
The anealing tests in the LXe detector were performed before each of the 2019 and 2020 runs.
During the tests, the detector was filled with gaseous xenon at room temperature.
The bias voltage was applied to each MPPC by a dedicated module which can supply sufficient current ($>20\,\mathrm{mA}$), and the LEDs in the LXe detector were used as a light source.
Seven MPPCs in 2019 and eight MPPCs in 2020 were annealed one by one at different bias voltages for different duration.
The annealing condition in 2020 is summarized in Table.~\ref{tab:mppc_anneal_cond}.

% VUV PDE
The $\PDEVUV$ light was measured after the annealing was finished and the detector was filled with LXe again because it was difficult to measure the $\PDEVUV$ in gaseous xenon at room temperature due to the high dark current.
A significant recovery of the $\PDEVUV$ was observed for the annealed MPPCs, as shown in Fig.~\ref{fig:annealing_2020}. The $\PDEVUV$ of neighboring MPPCs was also recovered by the heat transmitted from the annealed MPPCs.
The amount of the recovery (a ratio of the $\PDEVUV$ after to before the annealing) was correlated with the intensity of annealing: the induced current and duration.
The $\PDEVUV$ after annealing was 20\% for the intensely annealed MPPCs, which indicates the sufficient annealing can recover the $\PDEVUV$ fully.

\begin{figure}
    \centering
    \includegraphics[width=\hsize]{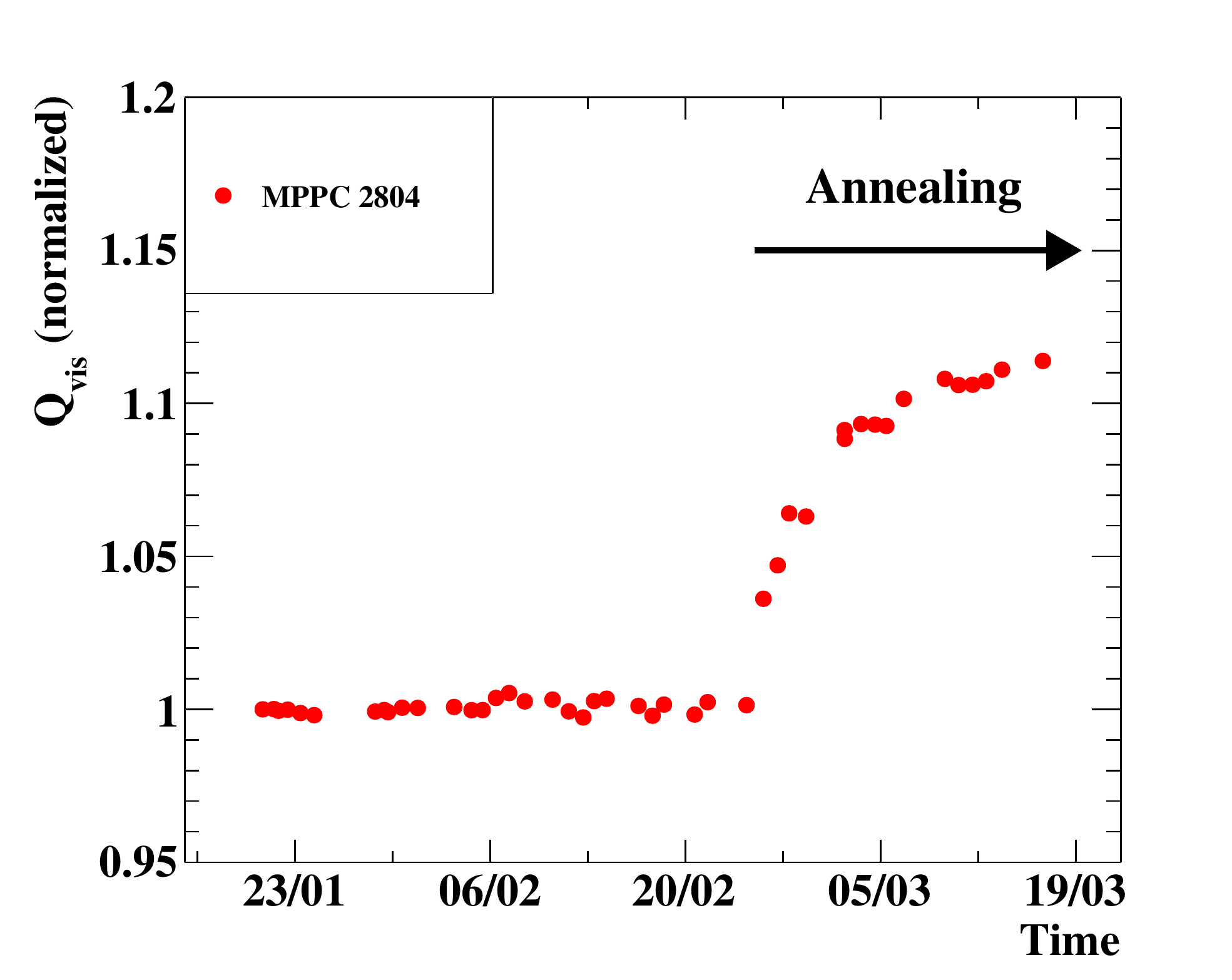}
    \caption{Time dependence of MPPC charge ($Q_{vis}$) for constant illumination of LED light during thermal annealing. 
    }
    \label{fig:annealing_qvis}
\end{figure}

% visible light
Interestingly, the $\PDEvis$ was also increased for the annealed MPPCs.
The relative increase was 0--10\% with respect to the $\PDEvis$ before the annealing.
As shown in Fig.~\ref{fig:annealing_visvuvcor}, the recovery of the $\PDEvis$ was well correlated with the recovery of the $\PDEVUV$.
This can be explained by the hypothesis that a fraction of incident visible light generates an electron-hole pair near the surface of the silicon bulk, and the restoration of the surface state positively impacts not only the VUV light detection but also the visible light detection.
The recovery of $\PDEvis$ was about 10\% of the recovery of $\PDEVUV$.
This ratio is consistent with the ratio between the decrease of $\PDEvis$ and $\PDEVUV$ under muon beam.
At the second annealing test in 2020, the recovery of $Q_{\text{vis}}$ (charge under constant illumination of LED light) was measured to monitor the progress of the annealing as in Fig.~\ref{fig:annealing_qvis}.
The annealing was stopped when the $Q_{\text{vis}}$ for the directly annealed MPPC was saturated.

\begin{table}[htb]
    \centering
    \caption{Configuration of annealing in 2020}
    \label{tab:mppc_anneal_cond}
    \begin{tabular}{cccc}
        \hline
        MPPC ID & current ($\mathrm{mA}$)  & voltage ($\mathrm{V}$)   & time (hours)\\
        \hline
        2763    & $15$--$21$ & $67$--$69$ & 488   \\
        2716    & $15$--$21$ & $68$       & 215.3 \\
        2807    & $22$       & $69$       & 125.5 \\
        2718    & $20$       & $67$       & 68.5  \\
        2719    & $22$       & $69$       & 84.3  \\
        2804    & $20$--$22$ & $68$--$69$ & 302.3 \\
        2674    & $20$       & $67$	       & 61    \\
        2675    & $21$       & $68$--$69$ & 157    \\
        \hline
    \end{tabular}
\end{table}

%I-V curve?

\section{Conclusion}\label{Conclusion}
This article reported a study on the radiation damage of VUV-sensitive MPPCs in the LXe detector for the MEG II experiment.
A significant degradation in the PDE was observed though the radiation dose level was well below the level at which standard MPPCs are affected.
The degradation of PDE for the visible light was much smaller than that for the VUV light.
These observations indicate that this degradation is a consequence of the surface damage.
It was found that the original PDE can be restored by thermal annealing.
%Thermal annealing was tested for several degraded MPPCs, and a significant recovery of the PDE was observed.
The study to identify the source of the radiation damage is in progress in a lab.

\section*{Acknowledgement}
We thank Paul Scherrer Institute as host laboratory.
%We also thank Hamamatsu Photonics K.K for their support.
This work was supported by JSPS KAKENHI Grant Numbers
JP25247034, % LXe
JP26000004, % Tokubetsu Suishin
JP20H00154, % Kiban A
JP19J21730, % Kobayashi
JP17J03308, % Ogawa DC2
and JSPS Core-to-Core Program, A. Advanced Reserach Networks JPJSCCA20180004.
\appendix
\section{Measurement of gain \& excess charge factor}\label{LEDCalib}
\begin{figure}
    \centering
    \includegraphics[width=\hsize]{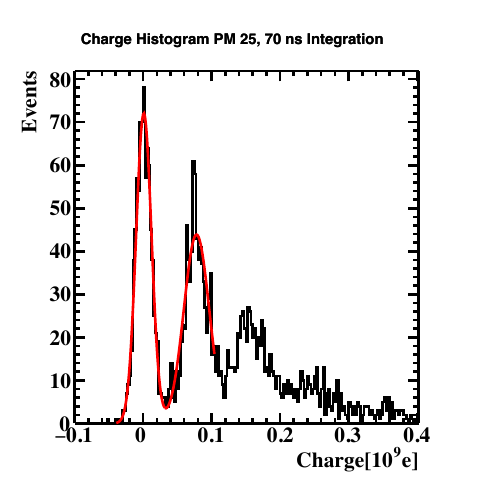}
    \caption{Charge spectrum of the MPPC for a weak LED light. The MPPC gain was obtained from the charge between two peaks corresponding to the zero photoelectron and the single photoelectron.}
    \label{fig:WeakLEDQHisto}
\end{figure}

The gain and excess charge factor ($F_{\text{EC}}$) are measured using LED light.
The intensity of the LED light is adjusted so that each MPPC detects only a few photons on average.
The MPPC gain is directly measured as the single photoelectron charge peak in the charge spectrum as shown in Fig.~\ref{fig:WeakLEDQHisto}.

The ECF can be extracted from the charge spectrum assuming the detected number of photons from the LED follows Poisson statistics.
The expected number of detected photo-electrons $\lambda$ is estimated using the fraction of zero-photoelectron events:
\begin{align}\label{exppe}
    \lambda = -\log (N_{0}/N_{\text{total}}).
\end{align}
The ECF can be measured as a ratio of the expected to measured mean charge as in Eq.~\ref{ECFeq}. The expected charge is the product of the measured gain and the expected number of photoelectrons.
\begin{align}\label{ECFeq}
    F_{\text{EC}} & = \overline{Q}_{\text{measured}}/Q_{\text{expected}} \\
        & = \overline{Q}_{\text{measured}}/(G\lambda). \label{Qexp}
\end{align}

\bibliographystyle{elsarticle-num-short}
\bibliography{VUVpaper.bib}

\end{document}